# Anisotropic Light-Matter Interactions in Single Crystal Topological Insulator Bismuth Selenide

Divya Rawat, Aditya Singh, Niraj Kumar Singh and Ajay Soni[*]

*School of Physical Sciences, Indian Institute of Technology Mandi, Mandi, 175005, HP India*

*Author to whom correspondence should be addressed: ajay@iitmandi.ac.in

Anisotropy of light-matter interactions in materials give remarkable information about the phonons and their interactions with electrons. We report the angle-resolved polarized Raman spectroscopy of single-crystal of $Bi_2Se_3$ to obtain the elements of Raman tensor for understanding the strength of polarization along different crystallographic orientations. Intensity variation in the polar plots corresponding to $E_g^1$ ~ 37 cm$^{-1}$, $A_{1g}^1$ ~71 cm$^{-1}$, $E_g^2$ ~ 130 cm$^{-1}$, and $A_{1g}^2$ ~ 173 cm$^{-1}$ suggests the higher differential polarizability along cross-plane (*bc*-plane). The polar patterns and the differences in elements of the Raman tensor provides the evidence of the fundamental electron-phonon and anisotropic light matter interactions in $Bi_2Se_3$.

## I. INTRODUCTION

Light-matter interaction helps to understand the many body physics and fundamentals of the electron and phonon coupling in materials.[1,2] Exploring the optical properties can provide significant understanding of the (*an*)-isotropic interaction of light along with the electronic susceptibility and permittivity (dielectric constant) of the materials. [3,4] Generally, the electric field vector ($\vec{E}$) of the incident and the scattered light are related through a complex matrix, known as Raman tensor ($T$) associated with the polarizability (α) of materials along three crystallographic orientations.[5] Recently, several layered materials such as $MoS_2$ [6], $WS_2$ , $MoSe_2$ [5], $PdTe_2$ [7] have been studied using Raman spectroscopy by controlling the polarization vector of incident and scattered light, to understand the dynamics of phonons along the different orientation of the crystal. Layered chalcogenide materials have been known for their anisotropic carrier relaxation times, which mainly arises due to their intriguing crystal structures and inherent anharmonicity.[8,9] Additionally, the Raman studies on ternary chalcogenides, $Bi_2GeTe_4$, $Sb_2SnTe_4$ have shown that electronic topological properties can also be coupled with phonons, which has been shown by the anomalous thermal behaviour of the Raman modes associated with bonds involved heavy elements. [8] Though several chalcogenide quantum materials have been explored extensively for their exotic electronic phenomena such as Shubnikov-de Haas quantum oscillations, [10] weak (anti)localization [11], thermoelectricity, superconductivity, charge-density waves and topological quantum insulating properties, yet the coupling of their topological electrons with phonons is less explored. [12-14] $Bi_2Se_3$ is one of the layered chalcogenides which has a fascinating layered crystal structure of five atoms (quintuple layers) stacked with van der Waals (vdWs) gaps and a crystal unit cell is composed of three quintuple layers. [15] Primarily, the topological studies on $Bi_2Se_3$ has a focus on investigating surface and bulk electronic structures using magneto-transport and angle-resolved photoemission spectroscopy studies, phonon dispersion, [16-19], but there are imperceptible reports on the anisotropic response of the inelastic light scattering. Since the topological quantum phenomena are associated with electrons, electron-phonon and electron-photon interactions [3,20], thus the investigation of the anisotropy of the electron-phonon-photon interaction, dynamics of phonon and evaluation of Raman-tensor are very important to explore. In this regard, the polarized Raman spectroscopy can provide a significant information about the light sensitive responses of single crystals along various orientations by controlling the polarization of both the incident and scattered photons to acquire the evidences of electron-phonon interactions and anisotropic behaviour. [21] In this work, we have discussed the angle resolved polarized Raman spectroscopy (APRS) to corroborate the interaction between the polarized light ($k_i$) and the crystallographic orientation of the single crystal $Bi_2Se_3$. The isotropic and anisotropic behaviour of phonons are studied with the rotation of crystal along two different configurations in *ab*-plane ($k_i||c$-axis) and *bc*-plane ($k_i||a$-axis), respectively. The observed anisotropic behaviour and polarizability of in-plane ($E_g$) and out-of-plane ($A_{1g}$) modes are quantified from the Raman tensor's elements. Our results open the opportunities to understand the role of anisotropic light-matter and electron-phonon interactions by both classical as well as quantum treatment of the Raman tensors obtained from the APRS analysis. The experimental details of synthesis and characterization of the single crystal are mentioned in supplemental materials.[22]

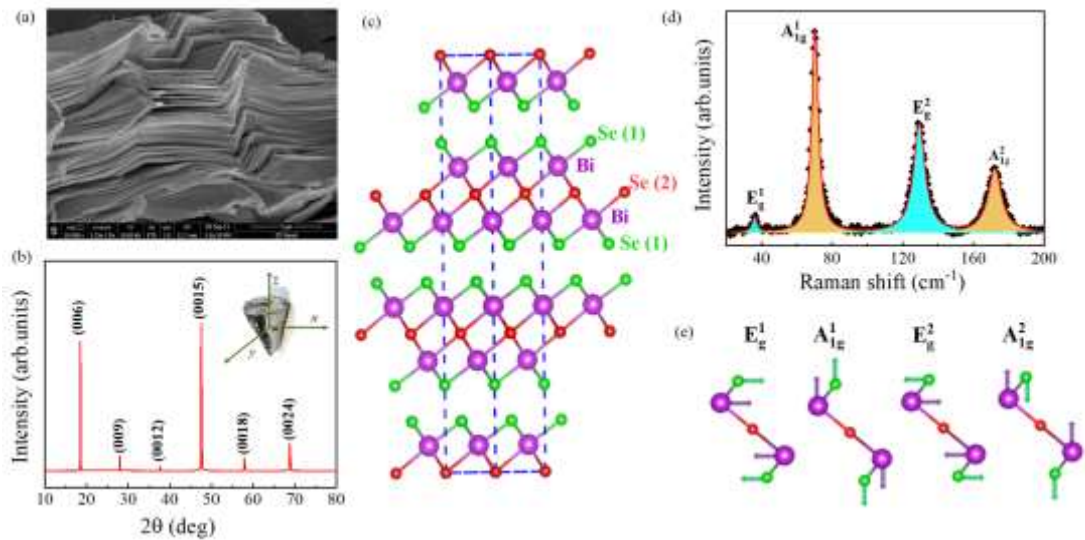

FIG. 1. (a) Electron microscopy image of the fractured cross section of layered $Bi_2Se_3$, (b) Powder X-ray diffraction pattern of single crystal showing the typical orientation along the *c*-axis, (inset: photograph of the grown sample), (c) Schematic of the crystal structure comprises of quintuple layers stacked with a weak Van der Waals gap, (d) Normalized Raman spectra and (e) Schematic of the atomic displacements of the $E_g$, and $A_{1g}$ modes.

The layered nature of the grown $Bi_2Se_3$ is shown in FESEM image (Fig. 1 (a)) and the XRD pattern in Fig. 1 (b), which confirms the orientation of the grown sample along *c*-axis.[23] Rietveld refinement of the XRD pattern of powdered $Bi_2Se_3$ provides the lattice parameters $a = b \sim 4.13$ Å, $c \sim 28.63$ Å, and unit cell volume ($V$) ~ 425 Å$^3$, (Fig. S1 of supplemental materials [22]). The residual resistance ratio (RRR ~ 2.11) has been evaluated from the low temperature resistance measurement (Fig. S2 of supplemental materials [22]), which shows a generate electron transport in a high quality of single crystal. [22] $Bi_2Se_3$ crystallizes in a rhombohedral crystal structure with space group R$\bar{3}$m (166), which is comprised of quintuple layers (*Se$^I$-Bi-Se$^{II}$-Bi-Se$^I$*) separated by weak vdW gap represented in Fig. 1(c). Here, *Se$^I$* and *Se$^{II}$* represents the different chemical environment of Se atoms in the unit cell. [24,25] The primitive unit cell of Bi$_2$Se$_3$ has fifteen zone-center vibrational modes, three acoustic and twelve optical, which can be represented by: $\Gamma = 2E_g + 2A_{1g} + 2E_u + 2A_{1u}$, where $A_{1g}$ and doubly degenerate $E_g$ are Raman active modes, whereas $2A_{1u}, 2E_u$ are the infra-red active modes.[24] The normalized room temperature Raman spectra, having modes at ~ 37 cm$^{-1}$ ($E_g^1$), ~ 71 cm$^{-1}$ ($A_{1g}^1$), ~ 130 cm$^{-1}$ ($E_g^2$), and ~ 173 cm$^{-1}$ ($A_{1g}^2$), is shown in Fig. 1(d) and the corresponding schematics of atomic displacements are illustrated in Fig. 1(e). The modes $A_{1g}^1$ ($A_{1g}^2$) and $E_g^1$ ($E_g^2$) have a different polarizability as they involve the out-of-plane and in-plane displacements in symmetric (anti-symmetric) stretching, respectively. Thus, angle-resolved polarized spectra (APRS) is an important tool to provide the detailed information on the interaction of the light along the different orientations of the crystal for estimation of elements of Raman tensor.

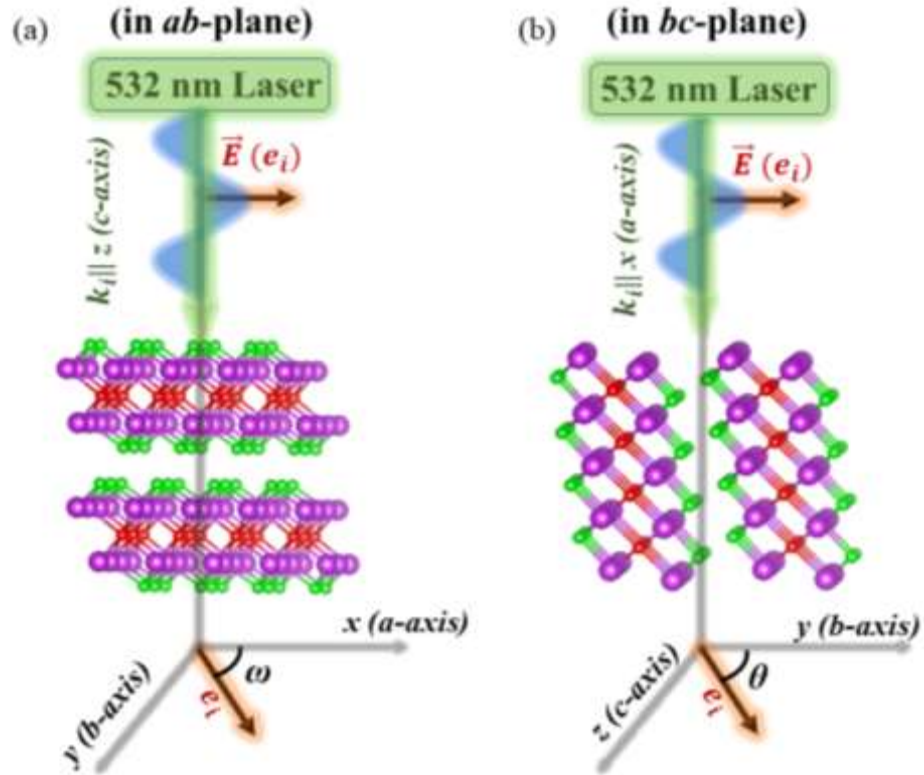

FIG. 2. Schematic representation of the two configurations used for APRS studies on Bi$_2$Se$_3$ crystal, where polarized laser ($k_i$) incidents along (a) *c*-axis (on *ab*-plane) and (b) normal to *c*-axis (*bc* -plane). Here, $\omega$ and $\theta$ correspond to the angle between electric polarization vector ($e_i$) of incident light with *a*-axis (in *ab*-plane) and *b*-axis (in *bc*-plane), respectively.

Fig. 2 represents the two configurations used for the APRS measurements, where crystallographic axes *a, b,* and *c* are taken as equivalent to *x, y,* and *z* axes of rotating stage. For the first configuration (Fig. 2 (a)), the incident laser ($k_i$) is parallel to the *c*-axis and electric polarization vector ($e_i$) is making an angle *ω* with the *a*-axis (in *ab*-plane). Hence, the scattering configuration is defined as $z(xx)\bar{z}$, and the corresponding polarization vector of incident and scattered light are $\vec{e_i} = \vec{e_s} =$ *(cos ω, sin ω, 0)*. For the second configuration (Fig. 2 (b)), the incident laser ($k_i$) is parallel to *a*-axis and electric polarization vector ($e_i$) is making an angle *θ* with the *b*-axis (in *bc*-plane). Correspondingly, the scattering configuration is defined as $x(yy)\bar{x}$ and the polarization vector of incident and scattered light are $\vec{e_i} = \vec{e_s} =$ *(0, cos θ, sin θ)*. Being isotropic in *ab*-plane, $Bi_2Se_3$ crystal does not have any changes in intensity along *a* and *b* axes while the anisotropic light-matter interactions along *c* axis and the details of Raman tensor is not reported in the literature.

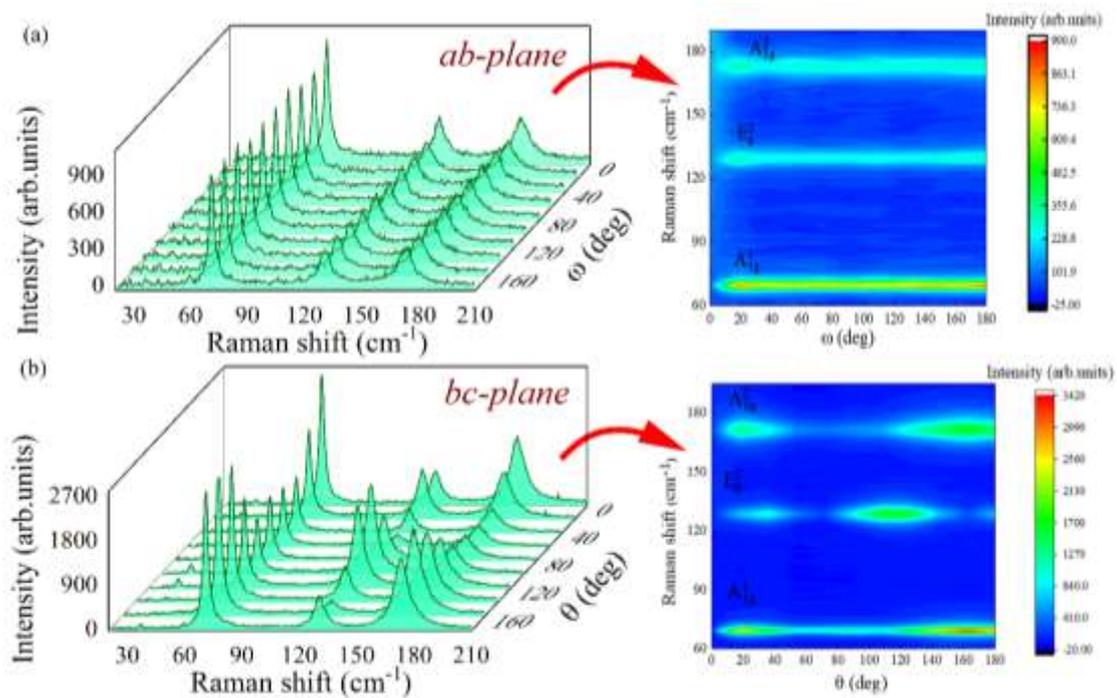

FIG. 3. Angle dependent polarized Raman spectra (a-b) and corresponding polarized Raman colour plot with the rotation of the $Bi_2Se_3$ sample in parallel configuration of polarized incident ($e_i$) and scattered ($e_s$) light along *ab* as well as *bc*-plane. Colour scale on the right side shows the intensity variation of Raman modes.

Polarized Raman spectra with the rotation of crystal along both *ab*(/*bc*)-plane and corresponding colour plot is shown in Fig. 3. The intensity of $A_{1g}^1$ ($A_{1g}^2$) and $E_g^1$ ($E_g^2$) modes are not changing along *ab*-plane (Fig. 3 (a)), whereas a periodic alteration has been observed along *bc*- plane (Fig. 3 (b)). The results indicate that there is an existence of anisotropy along the *bc*-plane as compared to *ab*-plane, which can be examined clearly from polar plots. According to classical treatment of Raman tensor, the inelastic process can be explained by the scattering from an extended medium, where the variations of the polarization can be expressed as a derivative of the susceptibility of the materials.[21] The contribution of such spatial symmetry to the Raman scattering intensity ($I$) can be expressed as $\langle e_i|Ṭ|e_s\rangle^2$, where $Ṭ$ is the Raman tensor for a given mode. [24] Thus, the elements of Raman tensor of $A_{1g}$ and double degenerate $E_g$ modes can be represented as:

$$Ṭ(A_{1g}) = \begin{bmatrix} \eta e^{i\phi_\eta} & 0 & 0 \\ 0 & \eta e^{i\phi_\eta} & 0 \\ 0 & 0 & \beta e^{i\phi_\beta} \end{bmatrix},$$

$$Ṭ(E_g) = \begin{bmatrix} \gamma e^{i\phi_\gamma} & 0 & 0 \\ 0 & -\gamma e^{i\phi_\gamma} & \delta e^{i\phi_\delta} \\ 0 & \delta e^{i\phi_\delta} & 0 \end{bmatrix} ; \begin{bmatrix} 0 & -\gamma e^{i\phi_\gamma} & -\delta e^{i\phi_\delta} \\ -\gamma e^{i\phi_\gamma} & 0 & 0 \\ -\delta e^{i\phi_\delta} & 0 & 0 \end{bmatrix},$$

Here the values corresponding to $\eta$, $\beta$, $\gamma$, and $\delta$ indicate the amplitudes whereas $\phi_\eta$, $\phi_\beta$, $\phi_\gamma$, and $\phi_\delta$ are the complex phases of the elements of Raman tensor. [21] Additionally, the magnitude of each tensor element is related with the specific mode and the crystal symmetry of the material. The calculated intensities for the estimation of the $Ṭ(E_g)$ has contributions from both the doubly degenerate $E_g$ modes, thus added altogether. Using the Raman selection rule, $|\langle e_i|Ṭ^*|e_s\rangle|^2$, under both *ab*(/*bc*)-plane, the scattering intensity of all modes have been calculated (Table I), which clearly showed the distinct strength of interaction of polarized light with the crystal's axes. [5,6,26,27]

TABLE I. Mathematically derived intensity of modes using Raman selection rules.

| S.no. | Configuration | Raman scattering intensity |
|---|---|---|
| 1. | *ab*-plane | $I^{\|}_{A_{1g}}(k_i//c\text{-axis}) = |\eta|^2$ |
|  |  | $I^{\|}_{E_g}(k_i//c\text{-axis}) = |\gamma|^2$ |
| 2. | *bc*-plane | $I^{\|}_{A_{1g}}(k_i//a\text{-axis}) = |\eta|^2 sin^4\theta + |\beta|^2 cos^4\theta + \frac{1}{2}|\eta||\beta|sin^2(2\theta)cos_{\varphi_{\eta\beta}}$ |
|  |  | $I^{\|}_{E_g}(k_i//a\text{-axis}) = |\gamma|^2 cos^4\theta + |\delta|^2 sin^2 2\theta - |\delta||\gamma|\sin(2\theta)cos^2\theta \times 2cos_{\varphi_{\gamma\delta}}$ |

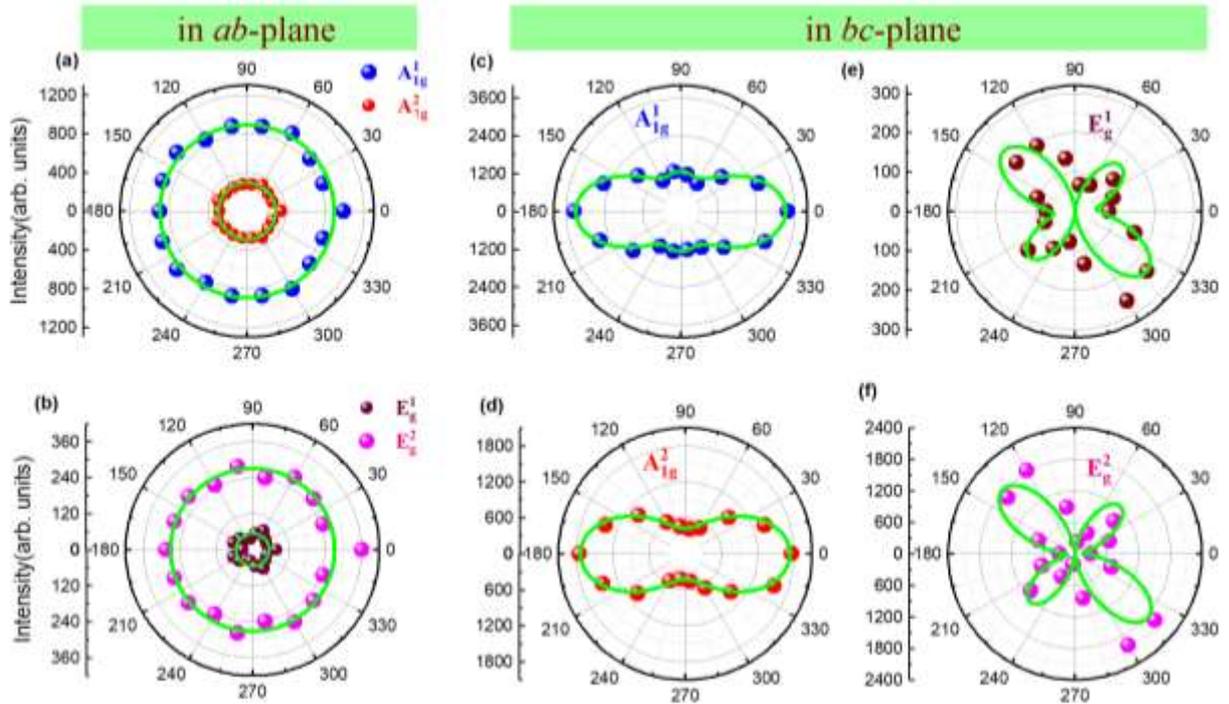

FIG. 4. Intensities of polar plots for $A_{1g}^1$, $A_{1g}^2$, $E_g^1$, $E_g^2$ modes in *ab*-plane (a-b), and in *bc*-plane (c-f). Here, solid symbols and green line represent the experimental data fitting of the data using equation in Table I, respectively.

Further, the understanding of the isotropic behaviour along *ab*-plane of the intensity of $A_{1g}$ and $E_g$ modes are depicted as circular shapes of the polar intensity plots (Fig. 4 (a-b)). On the other hand, the shape of polar plots for $A_{1g}$ (Fig. 4 (c-d)) and $E_g$ (Fig. 4 (e-f)) modes along *bc*-plane are different from *ab*-plane showing the anisotropy of the light matter interaction along crystallographic orientation. The intensities of all modes are stronger along *bc*-plane in comparison to the *ab*-plane, which advocates the higher differential polarizability along *bc*-plane. Similar observations on the anisotropic light-matter interaction in *bc*-plane have been reported for Graphene, hBN, 2H- MoSe$_2$, MoS$_2$.[5,6,28] Fascinatingly, the out of plane modes at ~ 71 cm$^{-1}$ and ~ 173 cm$^{-1}$, (Fig. 4 (c-d)), have $A_{1g}$ symmetry but showing considerably different polar pattern at 90° and 270° rotations. The anomalous polarization dependence of the Raman intensities appeared because of the difference in Raman scattering cross-section through the second-order susceptibility or the electron–phonon interactions.[21]

To understand the discrepancy, the microscopic quantum description of Raman tensor has been adopted, which involved the electron-phonon interaction in addition to the electron-photon. [29] Here, the total Raman intensity is described by the product of both the electron-photon and electron-phonon interactions. Hence, the Raman tensor ($Ʇ_{ij}^k$) associated with all modes can be given by:

$$Ʇ_{ij}^k = \frac{1}{V} \sum_{v,c,c\prime} \sum_{q\prime} \frac{\langle \Psi v(\vec{q}) | \vec{e}_s . \vec{\nabla} | \Psi_{c\prime}(\vec{q}) \rangle \langle \Psi_{c\prime}(\vec{q}) | H_{ep}^k | \Psi_c(\vec{q}) \rangle \langle \Psi_c(\vec{q}) | \vec{e}_i . \vec{\nabla} | \Psi_v(\vec{q}) \rangle}{(E_L - E_{cv}(\vec{q}) - i\Gamma_c)(E_L - \hbar\omega_{ph}^k - E_{c\prime v}(\vec{q}) - i\Gamma_{c\prime})}$$

Here, the numerator consists of the product of three matrix elements, (i) the electron-phonon (*e-ph*) matrix elements ($\langle \Psi_{c\prime}(\vec{q}) | H_{ep}^k | \Psi_c(\vec{q}) \rangle$) and two electron-photon matrix elements for incident and scattered light (ii) ($\langle \Psi_c(\vec{q}) | \vec{e}_i . \vec{\nabla} | \Psi_v(\vec{q}) \rangle$, (iii) $\langle \Psi v(\vec{q}) | \vec{e}_s . \vec{\nabla} | \Psi_{c\prime}(\vec{q}) \rangle$), where $\vec{e}_i$ and $\vec{e}_s$ are the polarization vectors of incident and scattered light, respectively.[29] The summation is over the electronics branches in conduction ($c, c\prime$) and valance ($v$) bands along with all wave vectors with first Brillouin zone. $\Gamma_c$ and $\Gamma_{c\prime}$ are the broadening factor associated with the lifetime of photo-excited states. The inclusion of *e-ph* matrix element gives the major differences among both the out of plane $A_{1g}$ modes. Thus, different patterns of polar plots for $A_{1g}^1$, and $A_{1g}^2$ modes indicate electron-phonon interactions in Bi$_2$Se$_3$, similar to the observations in other anisotropic layered chalcogenides like WS$_2$, ReS$_2$, GaTe, PdSe$_2$ and black phosphorus.[21,29-31] In contrast to the $A_{1g}$ modes, the polar plots of $E_g$ modes show four-lobbed polar pattern (Fig. 4 (e-f)) with the rotation of the crystal, which indicates the maximum strength of anisotropic nature in *bc*-plane. To understand the behaviour of polar plots related to $E_g$ modes, the spectra have been captured by controlling the polarization of incident light ($e_i$). This configuration is done by rotating half wave plate from 0° to 360° while keeping sample stage and analyzer fixed (Fig. S3 of supplemental materials. [22]) Here, the intensity of both $A_{1g}$ modes (Fig. S3 (a-b) of supplemental materials [22]) showed analogous polar pattern with polarization angle, whereas $E_g$ modes (Fig. S3 (c) of supplemental materials [22]) exhibited a low dependency on the rotation of the half wave plate. This discrepancy of the $E_g$ modes between the rotation of crystallographic axis and incident laser suggest the anisotropic behaviour along *bc*-plane. [5,6] Anisotropic light-matter interaction has been understood by estimating the amplitude and phase difference of Raman tensor's element, which mainly contain the information of differential polarizability along different orientation. To estimate the Raman tensor elements of all modes, we have fitted the experimental data (in Fig 4) using the intensity's expressions given in Table I and the obtained details are presented in Table II.

TABLE II. Estimated Raman tensor elements obtained from the fitting of experimental data (Fig 4).

| Modes | Raman tensor | |
|---|---|---|
| | *ab*-plane | *bc*-plane |
| $A_{1g}^1$ | $\begin{bmatrix} 30 & 0 & 0 \\ 0 & 30 & 0 \\ 0 & 0 & \beta \end{bmatrix}$ | $\begin{bmatrix} 35 & 0 & 0 \\ 0 & 35 & 0 \\ 0 & 0 & 57e^{i0.37\pi} \end{bmatrix}$ |
| $A_{1g}^2$ | $\begin{bmatrix} 17 & 0 & 0 \\ 0 & 17 & 0 \\ 0 & 0 & \beta \end{bmatrix}$ | $\begin{bmatrix} 21 & 0 & 0 \\ 0 & 21 & 0 \\ 0 & 0 & 41e^{i0.24\pi} \end{bmatrix}$ |
| $E_g^1$ | $\begin{bmatrix} 8 & -8 & \delta \\ -8 & -8 & \delta \\ \delta & \delta & 0 \end{bmatrix}$ | $\begin{bmatrix} 8 & -8 & -13e^{i0.39\pi} \\ -8 & -8 & 13e^{i0.39\pi} \\ -13e^{i0.39\pi} & 13e^{i0.39\pi} & 0 \end{bmatrix}$ |
| $E_g^2$ | $\begin{bmatrix} 16 & -16 & \delta \\ -16 & -16 & \delta \\ \delta & \delta & 0 \end{bmatrix}$ | $\begin{bmatrix} 14 & -14 & -38e^{i0.32\pi} \\ -14 & -14 & 38e^{i0.32\pi} \\ -38e^{i0.32\pi} & 38e^{i0.32\pi} & 0 \end{bmatrix}$ |

In *ab*-plane, all modes show isotropic behaviour (Fig 4a and 4b), hence for ҵ ($A_{1g}$) and ҵ ($E_g$), the component of Raman tensor, $\eta$ ($A_{1g}^1 \sim 30$ and $A_{1g}^2 \sim 17$) and $\gamma$ ($E_g^1 \sim 8$ and $E_g^2 \sim 16$), have been evaluated from the fitting of polar plots. As the propagation vector $k_i$ of incident light is along the *c*-axis, there is no polarization along *c*-axis, thus, $\beta$ for out of plane $A_{1g}$ mode is not evaluated while $\beta$ is zero for in-plane $E_g$ modes. Here, the phase factor ($\emptyset_\eta$) is zero due to isotropic responses in *ab*-plane. On the other hand, in *bc*-plane (Fig. 4c and 4d), the component of Raman tensor, ɳ ($A_{1g}^1 \sim 35$ and $A_{1g}^2 \sim 21$) and $\beta$ ($A_{1g}^1 \sim 57$ and $A_{1g}^2 \sim 41$) have been evaluated and the phase factor between $\eta$ and $\beta$ ($\emptyset_{\eta\beta}$) is $\sim 67.3°$ ($0.37\pi$) for ($A_{1g}^1$) and $\sim 44°$ ($0.24\pi$) for ($A_{1g}^2$), which is arising due to the anisotropic responses. Additionally, the elements of Raman tensor for in-plane modes are $\gamma$ ($E_g^1 \sim 8$ and $E_g^2 \sim 14$) and $\delta$ ($E_g^1 \sim 13$ and $E_g^2 \sim 38$) and the phase factor between $\gamma$ and $\delta$ ($\emptyset_{\gamma\delta}$) is $\sim 71°$ ($0.39\pi$) for ($E_g^1$) and $\sim 58.4°$ ($0.32\pi$) for ($E_g^2$). Overall, for out of plane $A_{1g}$ modes, $\beta > \eta$, ($57 > 35$ for $A_{1g}^1$ and $41 > 21$ for $A_{1g}^2$), which indicates that differential polarizability is significantly higher and anisotropic along *c*-axis (schematic Fig 1e). By comparing the tensor matrices of out of plane modes, it is clearly evident that symmetric stretching ($A_{1g}^1$) induces larger dipole moment (higher polarizability) than anti-symmetric stretching ($A_{1g}^2$) and the situation is completely otherwise for in-plane modes $E_g^1$ and $E_g^2$ as confirmed by the smaller magnitude of Raman tensor elements in Table II. For both the *ab*- and *bc*-plane, the comparison of relative magnitude of Raman tensor elements for of $A_{1g}^1$ ($|\eta_{bc-plane}/\eta_{ab-plane}| \sim 1.16$) and $E_g^2$ ($|\gamma_{bc-plane}/\gamma_{ab-plane}| \sim 1.14$), which authenticate the estimated elements of the Raman tensor. [6] Comparing the APRS estimated Raman tensor elements with studies on MoSe$_2$, MoS$_2$, WSe$_2$, PdTe$_2$, it is clear that the laser polarization dependence Raman spectra demonstrates the anisotropic light-matter interactions in Bi$_2$Se$_3$.

In Summary, the Raman tensor for all modes of single crystal Bi$_2$Se$_3$ corresponds to $E_g^1$ ~ 37 cm$^{-1}$, $A_{1g}^1$ ~70 cm$^{-1}$, $E_g^2$ ~ 129 cm$^{-1}$, and $A_{1g}^2$ ~ 172 cm$^{-1}$ have been systematically studied by APRS measurements along both *ab*(/*bc*)-plane under parallel polarization ($e_i \parallel e_s$) scattering configuration. We have estimated the amplitude and phase difference of the tensor elements by fitting the experimental results with the intensity expression obtained by applying Raman selection rule. The different shapes of polar plot of the similar vibrational symmetry ($A_{1g}$) represents the different interaction of electrons with phonons, which provide the evidence of electron-phonon coupling. Among two different orientations (*ab*(/*bc*)-plane) of single crystal, strong polarization dependence has been observed along *bc*-plane for both $A_{1g}$ and $E_g$ modes, which is showing the anisotropic light matter interaction in Bi$_2$Se$_3$.

## Acknowledgement

We would like to thank IIT Mandi for the instruments and research facilities. A.S would like to acknowledge DST-SERB for funding (Grant No. CRG/2018/002197).

# Supplemental Material

# Anisotropic Light-Matter Interactions in Single Crystal Topological Insulator Bismuth Selenide

Divya Rawat, Aditya Singh, Niraj Kumar Singh and Ajay Soni[*]

*School of Physical Sciences, Indian Institute of Technology Mandi, Mandi, 175005, HP India*

*Author to whom correspondence should be addressed: ajay@iitmandi.ac.in

In this supplemental file, we are providing the details of the synthesis, characterization techniques and selected data complementing the main text.

**(a) Synthesis and characterization details.**

Single crystal of $Bi_2Se_3$ was synthesized using dual zone vertical Bridgman furnace, by taking a stoichiometric amounts of bismuth ingot and selenium shots (both 99.999% pure) in a quartz ampoule, which was then vacuum sealed at $10^{-5}$ mbar. The ampoule was kept in a box furnace at 1123 K for 15 hr for homogenization followed by hanging it in Bridgman furnace. The temperature of the hot zone and cold zone were kept at 1003 K and 953 K, respectively. The translation rate of the motor for the vertical motion of quartz tube from hot zone to cold zone was fixed at 2 mm/hr. X-ray diffraction (XRD) was carried out using rotating anode Rigaku SmartLab diffractometer equipped with $CuK_\alpha$ radiation ($\lambda$ = 1.5406 Å) and in Bragg-Brentano geometry. Rietveld refinement of the Powder-XRD pattern was done to determine the crystal structure, lattice parameter, and phase purity. Resistance measurement was performed in the temperature range of 2 to 300 K using Quantum Design make physical properties measurement system (PPMS). Raman spectroscopy measurements were carried out using a Horiba LabRAM HR Evolution Raman spectrometer having 532 nm laser excitation, 1800 grooves/mm with the help of a Peltier cooled (CCD) detector. Ultra-low frequency filters were used to access low-frequency spectra, very close to laser line. To control the polarization state, a ($\lambda/2$) half-waveplate and an analyzer were used before objective lens and spectrometer to select the desired polarization component of the incident and scattered light, respectively. To study the light-matter interaction on the crystallographic axis of $Bi_2Se_3$, the sample was kept on the stage rotating from ~ 0º to ~ 360º with a step of ~ 20º. The linearly polarized laser was directed on the sample and the scattered radiation was collected to the detector in backscattering geometry.

**(b) Rietveld refinement analysis.**

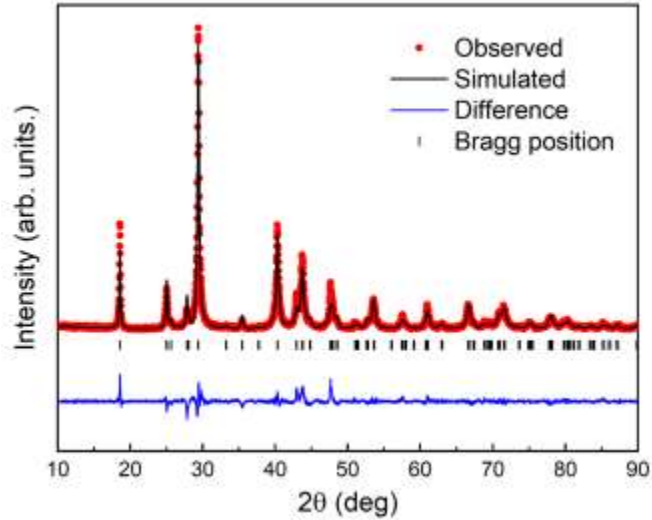

*FIG. S1:- Rietveld refined XRD pattern of single crystal $Bi_2Se_3$. Black closed circle represents the experimental data point, Solid red line represents the refined data.*

The as-synthesized $Bi_2Se_3$ crystal was ground into fine powder for XRD analysis. The phase purity of $Bi_2Se_3$ sample has been confirmed by Rietveld refinements of the powder XRD pattern. [1]The Fig. S1 shows the Rietveld refined XRD data. Goodness of fitting was showed by the extracting parameter, $\chi^2$ ~ 2.9.

**c) Resistance data of single crystal Bi$_2$Se$_3$:**

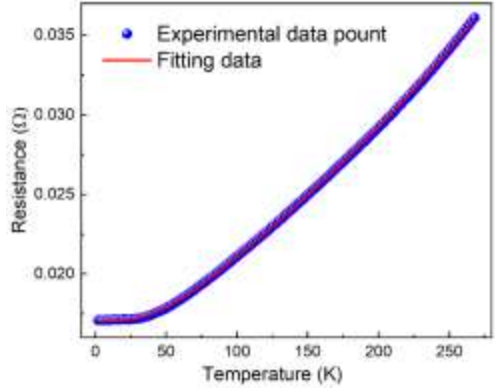

*FIG.S2:- Four-probe resistance measurement with the variation of the temperature.*

The electronic transport of the Bi$_2$Se$_3$ has been examined by the four probe resistance (R) and the temperature dependence is consistent with the behavior of degenerate semiconductors. The longitudinal resistance (R) is fitted using a phenomenological model: $R = R_0 + \lambda e^{-\theta/T} + \$T^2$, where the $\lambda$ and $\$$ appear for phonon scattering and electron-electron scattering, respectively.[2,3] The fitting parameter are evaluated and $\lambda \sim 12 \times 10^{-3}$ and $\$ \sim 1.74 \times 10^{-7}\ K^{-2}$, where smaller value of $\$$ suggests negligible electron-electron scattering in Bi$_2$Se$_3$. The residual resistance ratio (RRR ~ 2.11) shows a high quality of the single crystal.

**(d) APRS spectra of Bi$_2$Se$_3$ in bc-plane with the rotation of polarization vector of incident light while keeping the sample fixed.**

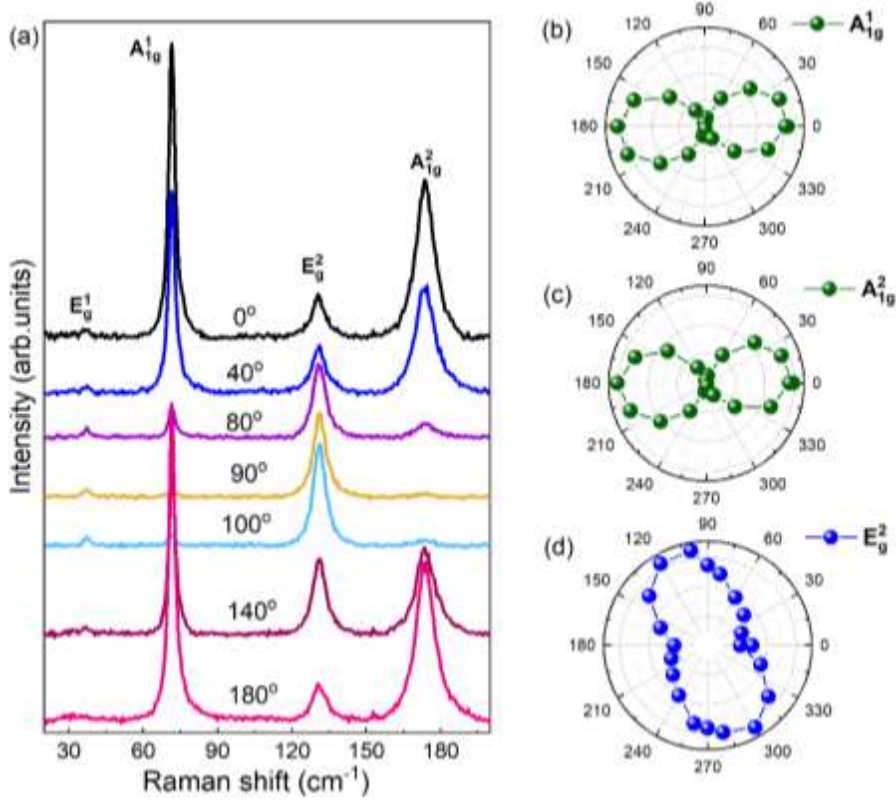

FIG. S3:- (a) APRS spectra and Polar plot of (b) $A_{1g}^1$ (c) $A_{1g}^2$ (d) $E_g^2$ of Bi$_2$Se$_3$ single crystal with the rotation of half-wave plate by keeping the sample fixed in bc-plane. Solid symbols represent the experimental data point.

APRS measurements has been performed in in parallel configuration ($e_i \parallel e_s$), where polarization vector of incident light has varied by rotating the half-wave plate, while keeping the stage of sample and analyzer fixed. Here, the intensity of both $A_{1g}$ modes showed expected two-lobed analogous polar pattern polar pattern with polarization angle. $E_g$ mode showed a low dependency on the rotation of the half wave plate, showed isotropic interaction on the rotation of polarization vector of incident light. [4]